\documentstyle[epsfig,array,aps]{revtex}
\begin{document}
\draft
\title{Cooling the Collective Motion of Trapped Ions to Initialize a Quantum
	Register\cite{disclaimer}}
\author{B.E.~King\cite{byline1}, C.S.~Wood, C.J.~Myatt, Q.A.~
Turchette, D.~Leibfried\cite{byline2}, W.M.~Itano, C.~Monroe, and
D.J.~Wineland}
\address{Time and Frequency Division, National Institute of Standards
and Technology, Boulder, Colorado 80303}
\date{\today}
\maketitle

\begin{abstract}
We report preparation in the ground state of collective modes of motion of  
two trapped $^9$Be$^+$ ions.  This is 
a crucial step towards realizing quantum logic gates which can entangle the 
ions' internal electronic states.  We find that heating of the modes 
of relative ion motion is substantially suppressed relative to that of the 
center-of-mass modes, suggesting the importance of these modes in future 
experiments.  
\end{abstract}

%\begin{multicols}{2}


In physics, quantum computation~\cite{steane:1998} 
provides a general framework for fundamental investigations into subjects 
such as entanglement, quantum measurement, and quantum information theory.
Since quantum computation relies on entanglement between qubits, any 
implementation of a quantum computer must offer isolation from the effects of 
decoherence, but also allow controllable and {\em coherent} interaction
between the qubits.  Cirac and Zoller~\cite{cirac-zoller:1995} have proposed 
an attractive scheme for realizing a quantum computer, which is scalable to an
arbitrary number of qubits.  Their scheme is based  
on a collection of trapped atomic ions, where each qubit (one per ion) is 
comprised of a pair of the ions' internal states, while quantum information 
is transferred between different ions using a particular quantized mode of 
the ions' collective motion.   This ``quantum data bus'' must first be 
initialized in a pure quantum state~\cite{cirac-zoller:1995}:  for
example, its ground state~\cite{footnote}.  The basics of this scheme have 
been demonstrated experimentally in a fundamental logic gate 
(a Controlled-NOT) operating between a motional mode of a single trapped ion 
and two of the ion's internal states~\cite{monroe:1995a}. 
In that work, the motional state was initialized in the ground state by laser 
cooling~\cite{monroe:1995}.  The next step towards implementing the 
Cirac-Zoller scheme is to cool at least one mode of collective motion of 
multiple ions to the ground state.  In this Letter, we describe the first 
experiments to realize this goal.  We also report significant difference
between the decoherence rates of the center-of-mass and 
non-center-of-mass modes of motion.

We confine $^{9}$Be$^{+}$ ions in a coaxial-resonator-based 
rf (Paul) trap, similar to that described in Ref.~\cite{jefferts:1995}.  
The electrodes in this trap are made from $125\,\mu$m-thick 
sheets of Be metal, as shown in Fig.~\ref{fig1}.  We apply a potential
$\phi(t)=V_0\cos(\Omega_{\mathrm{T}} t) + U_0$ to the (elliptical) ring 
electrode relative to the endcap electrodes.  If several ions are trapped and 
cooled, they will naturally align themselves along the major axis of the ring 
electrode.  The electrode's elliptical shape, in combination with $U_0>0$, 
allows a linear crystal to be maintained while suppressing rf-micromotion of 
the ions along this direction~\cite{myatt:1998}.  With 
$V_{0}\approx$ 520~V, $\Omega_{\mathrm{T}}/2 \pi \approx$ 238~MHz, and 
$U_0= $0~V, the pseudopotential oscillation frequencies are 
$(\omega_{x},\omega_{y},\omega_{z})/2 \pi \approx$ (4.6,12.7,17.0)~MHz.  With 
$U_0=18.2~\mathrm{V}$, the frequencies become (8.6,17.6,9.3)~MHz.  
Fig.~\ref{fig1} shows two ions confined in the trap and imaged with an $f/3$ 
lens system onto a position-sensitive photomultiplier tube.
 
The ions are cooled and probed with laser beams whose geometry is
indicated in Fig.~\ref{fig2}(a).  The relevant level structure of 
$^{9}$Be$^{+}$ is shown in Fig.~\ref{fig2}(b).  The quantization axis is 
defined by an applied static magnetic field; $|\bbox{B}|\approx 0.2$~mT.  The 
levels of interest for quantum logic operations are the 
$2s\:^2S_{1/2}|F=2,m_F=2\rangle$ and $2s\:^2S_{1/2}|F=1,m_F=1\rangle$ states, 
abbreviated  by $|\downarrow\rangle$ and $|\uparrow\rangle$, respectively.  
Laser beams D1, D2, and D3 are $\sigma^+$-polarized and focussed to nearly 
saturate the ions ($I_{sat}\approx$ 85~mW cm$^{-2}$).  Beams D1 and D2 
provide Doppler precooling in all three dimensions, and beam D3 prevents 
optical pumping to the $|F=2, m_{F}=1 \rangle $ state. The 
$|\downarrow\rangle \rightarrow 2p\:^{2}P_{3/2}|F=3,m_{F}=3 \rangle$ 
transition (radiative linewidth $\gamma/2\pi\approx 19.4$~MHz), 
driven by D2, is a cycling transition, which allows us to 
detect the ion's electronic state ($|\downarrow\rangle$ or $|\uparrow\rangle$) 
with nearly unit detection efficiency.  

Beams R1 ($\sigma^+/\sigma^-$-polarized) and R2 ($\pi$-polarized)
are used to drive stimulated Raman transitions between
$|\downarrow\rangle$ and $|\uparrow\rangle$,
through the virtual $2p\:^{2}P_{1/2}$ state~\cite{monroe:1995}.  
These beams are derived from a single laser, whose output is split by an 
acousto-optic modulator~\cite{thomas:1982}.  The beams are detuned by 
\hbox{$\Delta/2\pi \approx$ 40~GHz} to the red of the 
$2s\:^{2}S_{1/2} \rightarrow 2p\:^{2}P_{1/2}$ transition, and
their frequency difference is tuned around the $2s\:^{2}S_{1/2}$
hyperfine splitting of $\omega_{0}/2\pi \approx$ 1.25 GHz.  (Here, $\omega_0$ 
includes stable shifts of a few megahertz from the Zeeman and ac Stark 
effects.)  R2 is directed along 
$(-1/\sqrt{2})\bbox{\hat{x}}+(1/2)(-\bbox{\hat{y}}+\bbox{\hat{z}})$.  If 
R1$\perp$R2 as in Fig.~\ref{fig2}, then the Raman beam wavevector 
difference $\bbox{ \delta k}\parallel\bbox{\hat{x}}$, and the transitions are 
sensitive to ion motion only in this direction.  If, however, R1 is 
counterpropagating to R2, the transitions become sensitive to motion in all 
three dimensions.

When two cold ions are held in the trap and undergo small oscillations 
about their equilibrium positions, we may solve the equations of motion using 
normal mode coordinates. For two ions lying along the $x$-axis
there are two modes involving motion along this axis:  the center-of-mass 
(COM) mode (in which the ions move together with 
frequency $\omega_{\mathrm{COM}} = \omega_{x}$) and the stretch
mode (wherein the ions move out of phase, with frequency 
$\omega_{\mathrm{str}} = \sqrt{3}\omega_{\mathrm{COM}}$).  The other motional 
frequencies are $\omega_y$ ($y$ center-of-mass), $\omega_z$ 
($z$ center-of-mass), $\sqrt{\omega_y^2-\omega_x^2}$ ($xy$ rocking), and 
$\sqrt{\omega_z^2-\omega_x^2}$ ($xz$ rocking).

The lower traces in Fig.~\ref{fig3}, taken with 
$\bbox{ \delta k}\parallel\bbox{\hat{x}}$, shows an $x$-axis 
normal mode spectrum; results for the $y$- and $z$-modes are very similar.  We 
take the data with the following steps: first we turn on beams D1, D2, and D3 
for approximately 10 $\mu$s to Doppler cool the ions to the Lamb-Dicke regime, 
where the ions' confinement is much smaller than the laser wavelength.  Next, 
we turn off beam D2, and leave beams D1 and D3 on for 5~$\mu$s to optically 
pump both ions to the $|\downarrow\rangle$ state.  We then turn on only the 
Raman beams R1 and R2 for a time $t_{pr}$, with relative detuning 
$\omega_0 +\delta_{pr}$ (the ``Raman probe'' pulse).  Finally, we  drive the 
cycling transition with D2 and measure the ions' fluorescence.  We repeat the 
experiment at a rate of a few kilohertz while slowly sweeping $\delta_{pr}$.  
If the Raman beam difference frequency is resonant with a transition, then an 
ion is driven from $|\downarrow\rangle \rightarrow |\uparrow\rangle$ 
and the D2-driven fluorescence rate drops.

For a {\em single} ion, the carrier transition ($\delta_{pr}=0$) causes the 
population to undergo sinusoidal Rabi oscillations between 
$|\downarrow\rangle$ and $|\uparrow\rangle$~\cite{nonclass}. The 
effective Rabi frequency is $\Omega=g_1g_2/\Delta\approx 2\pi\times$~250~kHz, 
where $g_1$,~$g_2$ are the single-photon resonant Rabi frequencies of beams R1 
and R2. (We assume $\Delta\gg\gamma,\omega_m\gg\Omega$, where $\omega_m$ is
the frequency of the motional mode of interest.) If 
$\delta_{pr}=-\omega_x$ (the first lower $x$ sideband), then the transition 
couples the states $|\downarrow,n_x\rangle$ and $|\uparrow,n_x-1\rangle$, 
where $n_x$ is the vibrational level of the quantized motion along 
$\bbox{\hat{x}}$. In the Lamb-Dicke regime, the corresponding Rabi frequency 
is given by $\Omega_{n_x,n_x-1}=\eta_x\sqrt{n_x}\Omega$~\cite{nonclass}.  
Here, $\eta_x=x_0|\bbox{\delta k\cdot\hat{x}}|$ is the Lamb-Dicke parameter
(= 0.23 when $\omega_x/2\pi=$ 8.6~MHz) 
and $x_0=\sqrt{\hbar/(2m\omega_x)}$ is the spread of the $n_x=0$
wave function, with $m$ being the ion's mass).  (Note that if the 
ion is in the $n_x=0$ state of motion, this lower sideband vanishes.)
The first upper $x$ sideband transition ($\delta_{pr}=+\omega_x$) couples 
$|\downarrow,n_x\rangle$ and $|\uparrow,n_x+1\rangle$ with Rabi frequency 
$\Omega_{n_x,n_x+1}=\eta_x\sqrt{n_x+1}\Omega$.  

In the case of {\em two} ions driven on the carrier transition, each ion
independently undergoes Rabi oscillations between $|\downarrow\rangle$ and
$|\uparrow\rangle$ with Rabi frequency $\Omega$.  Since the laser beam waists  
($\approx 20\, \mu$m) are much larger than the ion-ion separation 
($\approx 2\, \mu$m), the ions are equally illuminated.  Nonetheless, if the 
micromotion of the two ions is different, then the reduction of the carrier 
(and sideband) transition strengths due to the micromotion will give a 
different Rabi frequency for each ion~\cite{myatt:1998,bible}.  This could be 
used as a means of selectively addressing the ions~\cite{turchette:1998};
however, in the present work the two ions' Rabi frequencies were equal. 

Since the sideband transitions affect the motional state, which is a shared 
property of both ions, such transitions produce entanglement between the ions'
spins and their collective motion~\cite{spin-sqz2}.  The system can no longer 
be treated as two, independent, two-level systems and the measured 
fluorescence following the Raman probe is a complicated function of the probe 
pulse duration $t_{pr}$.  For example, given an initial state 
$|\downarrow,\downarrow,n\rangle$ (where $n$ is the vibrational level of the 
COM or stretch motion along the $x$-axis) driven on the corresponding
lower sideband for a time $t_{pr}$, the wave function evolves as
%
%\end{multicols}
%
%\begin{picture}(0,0)
%\put (0,0){\line(1,0){240}}
%\put (240,0){\line(0,1){4.5}}
%\end{picture}
%
\begin{eqnarray}
  |\psi_n(t_{pr})\rangle=&&\left\{1-\frac{n}{2n-1}
			\bigl[1-\cos(G t_{pr})\bigr]
			\right\}|\downarrow,\downarrow,n\rangle
		 -i e^{i(\theta -\phi)/2}\sqrt{\frac{n}{2n-1}}\sin(G t_{pr})
			\frac{\Bigl(|\downarrow,\uparrow\rangle \pm
			e^{i \phi}|\uparrow,\downarrow\rangle\Bigr)
			|n-1\rangle}{\sqrt{2}}\nonumber\\
		    &&\mp e^{i\theta}\frac{\sqrt{n^2-n}}{2n-1}
			\bigl[1-\cos(G t_{pr})\bigr]
			|\uparrow,\uparrow,n-2\rangle\label{psi},
\label{eqn1}
\end{eqnarray}
%
%
%\begin{picture}(0,0)
%\put (260,0){\line(1,0){240}}
%\put (260,0){\line(0,-1){4.5}}
%\end{picture}
%
%\begin{multicols}{2}
%
\noindent where $G=\sqrt{2(2n-1)}\,\Omega\,\eta_{x,m}$ and 
$\theta$, $\phi$ are the sum and difference of the Raman beam phases at the
ions.  On the COM sideband (top sign in Eq.~(\ref{eqn1})), 
$\eta_{x,m}=\eta_{x,\mathrm{COM}}=\eta_{x}/\sqrt{2}$ (down by a
factor of $\sqrt{2}$ from the single-ion case due to the extra mass of the 
two-ion string), whereas on the stretch sideband (lower sign), 
$\eta_{x,m}=\eta_{x,\mathrm{str}}=\eta_{x}/\sqrt{2\sqrt{3}}$.  The expressions 
for transitions on the upper motional sidebands are similar.  If, before the 
Raman probe pulse, the ions have probability $p_n$ of being in the motional 
state $|n\rangle$, the subsequently-measured average fluorescence from the 
cycling transition is
%
%\end{multicols}
%
%\begin{picture}(0,0)
%\put (0,0){\line(1,0){240}}
%\put (240,0){\line(0,1){4.5}}
%\end{picture}
%
\begin{equation}
S(t_{pr})=\sum_{n}p_n\,\left(2\left|\langle\downarrow,\downarrow, n|
	\psi_n(t_{pr})\rangle\right|^2
	+\left|\langle\downarrow,\uparrow, n-1|\psi_n(t_{pr})\rangle
	\right|^2
	+\left|\langle\uparrow,\downarrow, n-1|\psi_n(t_{pr})
	\rangle\right|^2\right)\label{sig}.
\end{equation}
%
%\begin{picture}(0,0)
%\put (260,0){\line(1,0){240}}
%\put (260,0){\line(0,-1){4.5}}
%\end{picture}
%
%\begin{multicols}{2}
%
\noindent This signal is proportional to the expectation value of the number 
of atoms in the state $|\downarrow\rangle$.  For the data 
shown in Fig.~\ref{fig3}, $t_{pr}$ was chosen to maximize the sideband 
features.

The upper traces in Fig.~\ref{fig3} show the effects of adding several cycles
of Raman cooling~\cite{monroe:1995} on one particular $x$-mode after the 
Doppler cooling but before the probe pulse.  The reduction in the mean 
vibrational number $\langle n \rangle$ is indicated by the reduction in size 
of the lower sideband, which vanishes in the limit 
$\langle n\rangle\rightarrow~0$.  The data are consistent with a thermal state 
of $\langle n_{\mathrm{COM}} \rangle=0.11^{+0.17}_{-0.03}$ or 
$\langle n_{\mathrm{str}} \rangle = 0.01^{+0.08}_{-0.01}$.  This implies 
that the COM and stretch modes are in their ground states 
$90^{+3}_{-12}\%$ and $99^{+1}_{-7}\%$ of the time, respectively.  We have
also {\em simultaneously} cooled the COM and stretch modes along
$x$, to comparable values of $\langle n \rangle$ (and have separately
cooled the other four motional modes---$y$ and $z$ COM, $xy$ rocking, and $xz$
rocking---to near their ground states).

Each cycle of Raman cooling consists of: (1) a pulse of the Raman beams with 
their difference frequency tuned to one of the lower sidebands (COM
or stretch mode) and (2) optical repumping to the $|\downarrow\rangle$ state 
driven by beams D1 and D3.  The Raman transition reduces 
the vibrational energy by $\hbar\omega_m$ , whereas the repumping, 
on average, heats each mode by approximately the recoil energy 
($\ll \hbar\omega_m$).  Therefore, the ion is cooled 
through this process.  Five pulses of Raman cooling were used for the data
shown in Fig.~\ref{fig3}.  The exact durations of the Raman pulses were 
chosen to optimize the cooling rate---each pulse was approximately 
5~$\mu$s long.   

For an ion-trap implementation of a quantum computer, the motional modes are 
most susceptible to decoherence.  The ions' motional states lose coherence if 
they couple to (stochastic) electric fields caused by fluctuating potentials 
on the electrodes.  This leads to heating, which has previously been observed 
in single ions~\cite{monroe:1995,bible,diedrich:1989}; in 
Ref.~\cite{monroe:1995}, the heating drove the ion out of the motional (COM) 
ground state in approximately 1 ms.  We have performed similar heating 
measurements on the COM and non-COM modes of motion of two ions.  The results 
are summarized in Table~\ref{table1}.  The heating rate was determined by 
inserting a delay between laser cooling and the Raman probe.  The main results 
from these data are that the COM modes are heated out of the ground state much 
more quickly than the non-COM modes.  This can be explained as follows.

The COM modes, in which both ions move in phase, can be excited by a uniform
electric field.  However, no non-COM mode can be excited by a
uniform electric field~\cite{wineland:1975a}---
since these modes involve {\em differential} motion of the ions, they 
can only be driven by field gradients.  If the fluctuating
field at the ion (along the direction of motion of the mode of interest) is 
$E(t)$, an estimate of the corresponding field gradient is $E(t)/d$, where $d$ 
is a characteristic internal dimension of the trap.  For stochastic fields, 
the COM heating rate scales as $\langle E^2(t)\rangle$;  the non-COM mode 
heating rates scale as $\langle[\frac{E(t)}{d} \Delta x]^2\rangle$ 
(where $\Delta x$ is the ion-ion separation), down by a factor of $10^4$ for 
the present trap.  Similarly, other non-COM modes for more than two ions can 
only be excited by higher-order field gradients, leading to further reductions 
in their heating.

This suggests using non-COM modes for the quantum data bus in the 
Cirac-Zoller scheme.  Excitation of the ``spectator'' COM modes along 
the direction to which the Raman transitions are sensitive will still
alter the Rabi frequencies, but these effects will be higher order in 
the Lamb-Dicke parameter~\cite{bible}.  In the two-ion example, in the
Lamb-Dicke regime, the Rabi frequency for a first sideband transition 
$|n_1\rangle \rightarrow |n_1'\rangle$ on (cold) mode 1, given that 
(hotter) mode 2 is in the state $|n_2\rangle$, is~\cite{bible}
\begin{equation}
	\Omega_{n_1, n'_1} (n_2)
	= \Omega \eta_1 \sqrt{n_{1> \ }}e^{-(\eta_1^2 + \eta_2^2)/2}
	(1 - n_2 \eta_2^2),
\end{equation}
where $n_{1>}$ denotes the larger of $n'_1$ or $n_1$, and $\eta_1$ and $\eta_2$
are the Lamb-Dicke parameters for modes 1 and 2, respectively.  
Fluctuations in the Rabi frequency of mode 1 due to fluctuations in $n_2$ 
therefore occur in order $\eta_2^2$.  However, for the 
conditions of the present experiment, even if quantum logic operations were
performed using the $x$-stretch mode, the $x$-COM mode heating would still
limit the number of operations to around ten by the above mechanism.  Clearly,
this heating must be eliminated in future experiments.

The two-ion cooling results presented here are comparable to our previous
single-ion results~\cite{monroe:1995}, indicating that rf heating should not
be a concern for small numbers of ions~\cite{bible}.  Comparable cooling for
$N>2$ ions should not present any fundamental difficulties, as long as 
spurious overlaps of motional modes are avoided.

The preparation of a pure state of motion (the ground state) of multiple 
trapped ions represents the first step towards realizing quantum logic 
operations on them.  Such operations should lead to the creation of arbitrary 
entangled states of massive particles, including EPR- or GHZ-like spin 
states~\cite{EPR}.  Unlike other systems in which EPR states have been 
generated, it should be possible to reliably create these states on 
demand~\cite{turchette:1998} rather than by a selection process, and to detect 
them with nearly perfect efficiency~\cite{qj}.

We acknowledge support from the U.S. National Security Agency, Office of
Naval Research, and Army Research Office.  We thank Kristan Corwin,
David Kielpinski, and Matt Young for critical readings of the
manuscript.

\bibliographystyle{prsty}

\begin{thebibliography}{10}

\bibitem[*]{disclaimer}Work of the US government.  Not subject to US 
			copyright.

\bibitem[\dagger]{byline1}Electronic address:  kingb@ucsu.Colorado.EDU

\bibitem[\ddagger]{byline2}Present address:
Institut f\"{u}r Experimentalphysik, Universit\"{a}t Innsbruck, Innsbruck,
Austria.

\bibitem{steane:1998}
A. Ekert and R. Jozsa, Rev. Mod. Phys. {\bf 68},  733  (1996);
	A. Steane, Rep. Prog. Phys. {\bf 61}, 117 (1998).

\bibitem{cirac-zoller:1995}
J.~I. Cirac and P. Zoller, Phys. Rev. Lett. {\bf 74},  4091  (1995).

\bibitem{footnote}
If the ``data bus'' is in a thermal state with probability $P_0$ of being in
	the ground state then, if we ignore all other sources of error, a
	given quantum computation must be repeated $1/P_0$ times, on average,
	to give a correct answer.

\bibitem{monroe:1995a}
C. Monroe {\it et~al.}, Phys. Rev. Lett. {\bf 75},  4714  (1995).

\bibitem{monroe:1995}
C. Monroe {\it et~al.}, Phys. Rev. Lett. {\bf 75},  4011  (1995).

\bibitem{jefferts:1995}
S.~R. Jefferts, C. Monroe, E. Bell, and D.~J. Wineland, Phys. Rev. A 
	{\bf 51},  3112 (1995).

\bibitem{myatt:1998}
C.~J. Myatt {\it et~al.}, in ``Methods for Ultrasensitive Detection,''
	SPIE Proceedings Vol. 3270, edited by B.~L. Feary, (SPIE--International
	Society for Optical Engineering, Bellingham, WA, 1998), p. 131;
	R.~G. DeVoe, Phys. Rev. A {\bf 58} (to be published).

\bibitem{thomas:1982}
J.~E. Thomas {\it et~al.}, Phys. Rev. Lett. {\bf 48},  867  (1982).

\bibitem{nonclass}
D.~M. Meekhof {\it et~al.}, Phys. Rev. Lett. {\bf 76},  1796  (1996).

\bibitem{bible}
D.~J. Wineland {\it et~al.}, J. Res. Natl. Inst. Stand. Technol.
	{\bf 103}, 259 (1998) and e-print quant-ph/9710025;
	Forschr. Phys. {\bf 46}, 363 (1998).

\bibitem{turchette:1998}
Q.~A. Turchette {\it et~al.}, e-print quant-ph/9806012.

\bibitem{spin-sqz2}
D.~J. Wineland, J.~J. Bollinger, W.~M. Itano, and D.~J. Heinzen, Phys.
	Rev. A {\bf 50}, 67  (1994).

\bibitem{diedrich:1989}
F. Diedrich, J.~C. Bergquist, W.~M. Itano, and D.~J. Wineland, Phys.
	Rev. Lett. {\bf 42}, 403  (1989).

\bibitem{wineland:1975a}
D.~J. Wineland and H.~G. Dehmelt, J. Appl. Phys. {\bf 46},  919
  (1975); D.~F.~V. James, Phys. Rev. Lett. {\bf 81}, 317 (1998).

\bibitem{EPR}
A. Einstein, B. Podolsky, and N. Rosen, Phys. Rev. {\bf 47},  777  (1935);
   D.~M. Greenberger, M.~A. Horne, A. Shimony, and A. Zeilinger, Am. J.
   Phys. {\bf 58},  1131  (1990); M. Lamehi-Rachti and W. Mittig, Phys. 
   Rev. D {\bf 14}, 2543  (1976); E. Hagley, {\it et al.}, Phys. Rev. 
   Lett. {\bf 79}, 1  (1997).

\bibitem{qj}
R. Blatt and P. Zoller, Eur. J. Phys. {\bf 9}, 250 (1988).


\end{thebibliography}

%\end{multicols}
\begin{twocolumn}

\begin{figure}
\epsfig{file=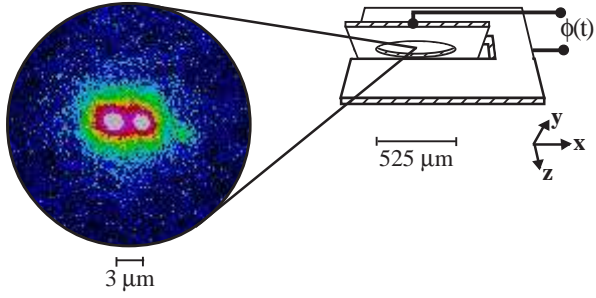}
\caption{Two ions trapped in an elliptical rf (Paul) trap.  The ring has 
	an aspect ratio of 3:2 and the major axis is 525~$\mu$m long.  The 
	slot which forms the endcaps is 250~$\mu$m across.  A potential
	$\phi(t)$ is applied to the ring (see text).  The Be 
	sheets are $\approx 125\mu$m thick.  With an $x$-axis pseudopotential 
	oscillation frequency $\omega_x /2\pi =$~4.6 MHz, the ion-ion spacing 
	is approximately 3~$\mu$m.}
\label{fig1}
\end{figure}

\begin{figure}
\epsfig{file=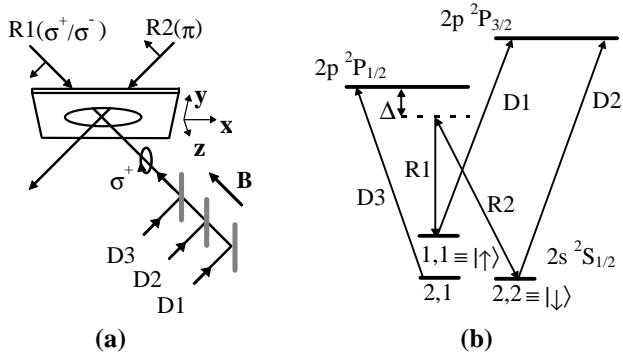}
\caption{ (a) Laser beam geometry.  The trap ring electrode is shown rotated 
	45$^\circ$ out of the page.  The endcap electrodes are omitted for
	clarity; see Fig.~\ref{fig1}.  A magnetic field $\bbox{B}$ of
	magnitude 0.2 mT defines the quantization axis along 
	$-(1/\protect\sqrt{2}) \bbox{\hat{x}} + 
	(1/2)(\bbox{\hat{y}} - \bbox{\hat{z}})$,
	and laser beam polarizations are indicated.  (b) Relevant
	$^9$Be$^+$ energy levels (not to scale), indicated by $F$, $m_F$ 
	quantum numbers in the ground state.  $^2P$ fine structure
	splitting is $\approx$~197 GHz, $^2S_{1/2}$ hyperfine splitting	is 
	$\omega_0 /2 \pi \approx$ 1.25 GHz, $2P_{1/2}$ hyperfine splitting is
	$\approx$~237 MHz, and the $^2P_{3/2}$ 
	hyperfine structure ($\ll\gamma/2\pi\approx19.4$~MHz) is not resolved. 
	All optical transitions are near $\lambda \approx$~313 nm, and 
	$\Delta /2 \pi \approx$~40 GHz. }
\label{fig2}
\end{figure}

\begin{figure}
\epsfig{file=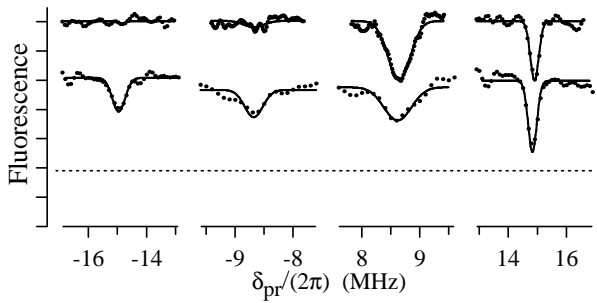}
\caption{Spectrum of sidebands due to two-ion $x$-axis normal mode motion:  
	(from left to right) lower stretch, lower COM, 	upper COM, and upper 
	stretch.  The ordinate 	is the detuning of the Raman probe beam 
	difference frequency from the carrier transition.  The abscissa shows 
	the ion fluorescence (proportional to the expectation value of the 
	number of atoms in the state $|\downarrow\rangle$), plus a constant 
	background (whose approximate level for the lower curves is indicated 
	by the dashed line).  The solid lines, meant as guides to the eye, are
	fits to Gaussians.  The lower traces show the effects of Doppler 
	cooling.  The upper traces, offset vertically for clarity, show the 
	effects of several pulses of Raman cooling on the mode which is 
	displayed.  Vanishing lower motional sidebands indicate cooling to the 
	ground state of motion.  The peak widths are consistent with the Raman 
	probe pulse lengths ($\approx 3\;\mu$s).}
\label{fig3}
\end{figure}

\begin{table}
\caption{Heating rates of the six normal modes of two trapped ions.  The Raman 
	beams were counterpropagating for the $y$- and $z$- axis data, making 
	the Raman probe	sensitive to motion in all three dimensions.  Note 
	that the COM modes are heated at a much higher rate than the non-COM 
	modes (see text).  (The precision with which the heating rates are 
	given for the last five modes is limited by measurement
	noise.)}
\label{table1}
\begin{tabular}{lcc}
	mode & $\omega_m/2\pi~\mathrm{(MHz)}$ 
	& $\delta\langle n\rangle/\delta t$~(ms$^{-1})$\\ \hline
	$x_{\mathrm{COM}}$ & 8.6 & $19^{+40}_{-13}$\\
	$y_{\mathrm{COM}}$ & 17.6 & $>10$\\
	$z_{\mathrm{COM}}$ & 9.3 & $>20$\\
	$x_{\mathrm{str}}$ & 14.9 & $< 0.18$\\
	$xy_{\mathrm{rocking}}$ & 15.4 & $< 1$\\ 
	$xz_{\mathrm{rocking}}$ & 3.6 & $< 0.5$
\end{tabular}
\end{table}

\end{twocolumn}

\end{document}